\documentclass[printer]{aa}
\usepackage{amssymb}
\usepackage{amsmath}
\usepackage{txfonts}
\usepackage{graphicx}
\usepackage{natbib}
\usepackage{rotating}
\usepackage{soul}
\usepackage{url}
\usepackage{epsfig}
\usepackage{longtable}

\usepackage{color}
\usepackage[utf8]{inputenc}

\begin{document}

\title{Reconstructing the projected gravitational potential of Abell 1689 from X-ray measurements}
\titlerunning{Gravitational potential of Abell 1689 from X-ray measurements}
\authorrunning{C. Tchernin et al.}

\author{Céline Tchernin\inst{1}, Charles L. Majer\inst{2,3}, Sven Meyer\inst{2}, Eleonora Sarli\inst{2}, Dominique Eckert\inst{1}, Matthias Bartelmann\inst{2}}
\institute{\inst{1}Astronomical Observatory of the University of Geneva, ch. d'Ecogia 16, 1290 Versoix, Switzerland,\\ e-mail: Celine.Tchernin@unige.ch,\\\inst{2}Universität Heidelberg, Zentrum für Astronomie, Institut für Theoretische Astrophysik, Philosophenweg 12, 69120 Heidelberg, \\\inst{3}Germany,Division of Medical Physics in Radiology, German Cancer Research Center (DKFZ), 69120 Heidelberg, Germany.}


\abstract
{Galaxy clusters can be used as cosmological probes, but to this end, they need to be thoroughly understood. Combining all cluster observables in a consistent way will help us to understand their global properties and their internal structure.}
{We provide proof of the concept that the projected gravitational potential of galaxy clusters can directly be reconstructed from X-ray observations.  We also show that this joint analysis can be used to locally test the validity of the equilibrium assumptions in galaxy clusters.}
{We used a newly developed reconstruction method, based on Richardson-Lucy deprojection, that allows reconstructing projected gravitational potentials of galaxy clusters directly from X-ray observations. We applied this algorithm to the well-studied cluster Abell~1689 and compared the gravitational potential reconstructed from X-ray observables to the potential obtained from gravitational lensing measurements. We also compared the X-ray deprojected profiles obtained by the Richardson-Lucy deprojection algorithm with the findings from the more conventional onion-peeling technique. }
{Assuming spherical symmetry and hydrostatic equilibrium, the potentials recovered from gravitational lensing and from X-ray emission agree very well beyond 500~kpc. Owing to the fact that the Richardson-Lucy deprojection algorithm allows deprojecting each line of sight independently, this result may indicate that non-gravitational effects and/or asphericity are strong in the central regions of the clusters.}
{We demonstrate the robustness of the potential reconstruction method based on the Richardson-Lucy deprojection algorithm and show that gravitational lensing and X-ray emission lead to consistent gravitational potentials. Our results illustrate the power of combining galaxy-cluster observables in a single, non-parametric, joint reconstruction of consistent cluster potentials that can be used to locally constrain the physical state of the gas.}

\maketitle

\section {Introduction}

The cold dark matter (CDM) model favored by substantial theoretical and observational evidence \citep[e.g., ][]{komatsu11} predicts the formation of small structures at early times and their later merging into larger systems. In the CDM framework, galaxy clusters are expected to be the latest gravitationally bound systems to form.

Galaxy clusters are mainly composed of three components: galaxies, hot gas, and dark matter. The dark matter is by definition impossible to observe directly, but it is expected to follow a universal density profile approximated by the Navarro-Frenk-White profile \citep[NFW,][]{NFW}. However, observations of the intracluster medium (ICM) and of cluster galaxies can constrain the dark-matter distribution itself. As the ICM fills the deep gravitational potentials, its distribution is governed by the depth and shape of the dark-matter potential well \citep[see, e.g.,][for a review]{ettori13}, hence the observables provided by the thermal plasma carry information on the gravitational potential.

The physical properties of the ICM can be directly studied in X-rays through thermal bremsstrahlung and line emission and at millimetrer wavelengths through the Sunyaev-Zel'dovich (SZ) effect. On the other hand, measurements of gravitational lensing in clusters constrain the gravitational tidal field and thus the curvature of the projected gravitational potential of these clusters. Our study is based on the idea that a simultaneous combination of as many galaxy-cluster observables as possible may lead to a consistent model for the projected cluster potential because under equilibrium assumptions, all these observables can be combined on the grounds of the gravitational potential.

In addition to improving the reconstruction of the projected cluster potential beyond gravitational lensing, combining as many cluster observables as possible will also allow us to cover a wider range of angular or radial scales. The joint analysis of weak and strong gravitational lensing allows us to map the dark-matter distribution within the entire cluster. Furthermore, because the emissivity depends on the square of the gas density, X-ray flux and temperature profiles are most reliable in and near the central region of the cluster \citep{ettori11}, while the thermal SZ effect has a shallower dependence on the local gas density because the thermal
SZ signal is proportional to the gas pressure integrated over
the line of sight \citep{planck13,eckert13}. Hence, owing to the different dependences of different observable signals on angular scales, weak-lensing and thermal-SZ measurements extend to relatively large distances from the cluster center, while strong-lensing and X-ray observations are more sensitive to regions near the cluster center. In combination, more reliable reconstructions of cluster potentials can be expected on all scales.

Another advantage of this joint analysis is that the assumptions made on the physical state of the gas can be verified. The lensing potential can be recovered without any equilibrium assumptions, while the gravitational potential reconstructed from X-ray and thermal-SZ observations is recovered assuming hydrostatic equilibrium. Therefore, the projected gravitational potential recovered from gravitational lensing observations can be compared with the gravitational
potential recovered from the hot ICM observations, and the result can be used to determine which clusters satisfy the assumptions and which do not. This will help us to distinguish the different processes occurring in clusters.

 We here focus on reconstructing the projected potential of galaxy clusters using X-ray observations. However, other observables may also be used, such as galaxy kinematics \citep{kinematics} or the thermal SZ effect \citep{SZ}. Our final goal is to set up one common likelihood function for the gravitational potential, joining information from all available observables. 
As a proof of concept, in this study the projected potential of the well-studied cluster Abell~1689 is reconstructed from X-ray observations. 

\paragraph{}
The cluster Abell~1689 is located at redshift $0.183$ in the Virgo constellation. As a well-known strong-lensing cluster, it has been the subject of many studies. The multiwavelength observations performed on this object extend from the mid-infrared, for instance, using \textit{ISOCAM} onboard the \textit{ISO} satellite \citep{fadda00}, to the X-ray band, for instance, using \textit{Suzaku} \citep{kawaharada10}, \textit{ROSAT} \citep{allen98}, \textit{Chandra}\footnote{see also the Chandra catalog of \citet{ACCEPT}.} \citep{xu10}, and \textit{XMM-Newton} \citep{andersson04,snowden08}. In visible light, the optical signal has been observed using the \textit{Subaru/Suprime-Cam} telescope and the \textit{Hubble Space telescope} \citep{broadhurst05}. This list is not comprehensive. 

The main result of these observations is that Abell~1689 is a dynamically active cluster: kinematics and X-ray studies indicate a merger aligned with the line of sight \citep[see, e.g.,][and references therein]{andersson04}. Combinations of different observations in joint analyses have been performed by many authors and have shown that the hydrostatic mass is lower than  the mass estimated from gravitational lensing measurements
in the central regions \citep[see, e.g.,][]{peng09, morandi11,miralda95, lokas06, sereno13}.

\citet{miralda95} combined strong-lensing and X-ray surface brightness measurements to test the physical state of the gas near the cluster center. The authors showed that the observed temperature in the central regions is lower than that expected from hydrostatic equilibrium. As a result, a non-negligible nonthermal pressure component should be present and act against gravity. According to these authors, this nonthermal pressure support could be associated with the merger process.
In a complementary analysis, \citet{lokas06} measured the velocity distribution of the galaxies in the field and identified several matter clumps aligned with the line of sight through the cluster, which do not interact with the cluster dynamically and might affect the lensing mass estimates without modifying the X-ray mass estimate. This provides an alternative explanation for the mass discrepancy observed between the two methods. 
In a different analysis of Abell~1689, \citet{peng09} combined weak-lensing with X-ray data from the \textit{Chandra} satellite. Considering the central part of the cluster out to $R\sim 1000~\mathrm{kpc/h,}$ they found that at radii larger than $R\sim200~\mathrm{kpc/h}$, the X-ray and lensing masses are consistent, whereas they found a discrepancy between the masses at smaller radii. These authors explained this discrepancy by a projection effect. This conclusion agrees with the results of \citet{sereno13}. In this study, the authors constrained the shape and orientation of the cluster in a joint analysis of the ICM distribution in the cluster. Using X-ray and SZ measurements, the cluster appears to have a triaxial shape elongated along the line of sight. According to these authors, this may explain the mass discrepancy. 

Finally, in a joint analysis of X-ray surface brightness with strong- and weak-lensing measurements, \citet{morandi11} inferred the fraction of nonthermal pressure required for the cluster to be in hydrostatic equilibrium, taking into account the triaxial shape of the cluster. The authors found that 20~\% of the total pressure should be nonthermal, most probably contributed by turbulent gas motion driven by hierarchical merger events. 
 
To summarize, a mass discrepancy has been observed in many studies, and different explanations have been suggested. To understand the origin of this discrepancy,  we here propose the following method: 1) reconstruct the three-dimensional gas distribution using a deprojection method that operates independently on each line of sight: the Richardson-Lucy deprojection algorithm \citep[R-L,][]{RL_lucy74,RL_lucy94};
and 2) substitute the mass with a local quantity, that is,\ the gravitational potential.

This procedure, applied here  to a joint analysis of X-ray and lensing observations, is expected to allow us to directly compare the dark matter and gas distributions and to test  the validity of the equilibrium assumptions at each projected radius. Using the gravitational potential should render the comparison with gravitational  lensing measurement more straightforward because the gravitational potential is directly observable from the lensing measurements.

To recover the cluster potential from X-ray measurements, we first deproject the X-ray data, then convert the X-ray three-dimensional profile into the three-dimensional gravitational potential using equilibrium assumptions, and project it again to compare it with the two-dimensional gravitational potential obtained by gravitational lensing. Applying this method with different geometries (spherical and triaxial) is expected to remove the degeneracy between the different possible explanations for the mass discrepancy. We test this algorithm here for the first time on a real cluster assuming spherical symmetry. The generalization to a spheroidal shape is ongoing.

This paper is structured as follows: In Sect.~\ref{sec:assum} we
list the assumptions of the method. In Sect.~\ref{test_RL}
we apply the R-L deprojection algorithm on the cluster Abell~1689 and compare this with the onion-peeling deprojection method \citep[O-P, see, e.g.,][for a detailed overview of this method]{OP_fabian81,kriss83}, which is the standard deprojection method in X-ray astronomy. In Sect.~\ref{sec:SphericalSym}, O-P and R-L deprojection methods are directly compared for different cluster observables, and in Sect.~\ref{sec:RL_OP} the advantage of the R-L method with respect to the O-P method is discussed. In Sect.~\ref{test_reconstruction}, the reconstructed projected gravitational potential obtained using the method developed in \citet{RL_method} is compared to the lensing potential of the cluster as recovered by J. Merten (private communication). Finally, the results are discussed in Sect.~\ref{sec:discussion}, and we conclude in Sect.~\ref{sec:conclusion}.


\section{Assumptions of the method}\label{sec:assum}

The general reconstruction formalism outlined in \citet{RL_method} is based on the following assumptions:
\begin{itemize}
\item The ICM is in hydrostatic equilibrium,
\begin{equation}\label{eq:eqHydro}
  \nabla P=-\rho \nabla\Phi\;,
\end{equation}
where $\Phi$ denotes the Newtonian gravitational potential, while $P$ and $\rho$ are the gas pressure and density.
\item The gas follows the polytropic relation
\begin{equation}\label{eq:eqPoly}
  \frac{P}{P_0}=\left(\frac{\rho}{\rho_0}\right)^{\gamma}\;,
\end{equation}
where the suffix $0$ denotes fiducial values for the pressure and for the gas density at an arbitrary radius $r_0$ and where $\gamma$ is the polytropic index.
\item The gas is ideal,
\begin{equation}\label{eq:idealGas}
  P=\frac{\rho}{\bar{m}} k_\mathrm{B}T\;,
\end{equation}
where $T$ is the gas temperature, $k_\mathrm{B}$ is Boltzmann's constant, and $\bar{m}$ is the mean mass of a gas particle.
\item The bremsstrahlung emission is assumed to dominate the X-ray observations. This assumption is justified as long as the temperature of the plasma does not fall below $\sim2\,\mathrm{keV}$ and the Fe~XXV-Fe~XXVI line complex at $6-7\,\mathrm{keV}$ is avoided \citep{peterson06}. The bolometric bremsstrahlung emissivity, $j_x$, can be written as
\begin{equation}\label{eq:Brem}
  j_x=CT^{1/2}\rho^2\;,
\end{equation}
where $C$ is an amplitude formed by all relevant physical constants. We integrated the emissivity over the energy band $[0.4-2]\,\mathrm{keV}$ ($j_{[0.4;2]}$) instead of over the entire electromagnetic spectrum. Introducing finite boundaries for the energy bands modifies the temperature dependence of the emissivity, but not its dependence on density. Since the density profile is steeper than the temperature profile, and the emissivity depends on the square of the density, the emissivity profile is dominated by the density profile and uncertainties in the temperature dependence can be considered negligible for now.
\item  Except for the natural constants and the averaged parameters, all quantities introduced here are assumed to depend on the cluster-centric radius. This dependence is not written explicitly to simplify the notation. The effective adiabatic index is assumed to be constant across the cluster. While this has been shown not to be true in general, it can be a good approximation as long as the core of the cluster is avoided \citep[see, e.g,.][and the discussion section]{tozzi01, capelo12, eckert13}.
\item The cluster is assumed to be spherically symmetric.
\end{itemize}

\section{Deprojection of the cluster Abell~1689} \label{test_RL}

To construct the surface-brightness profile of the cluster, we used an archival $13.4$ ks observation of Abell~1689 with \emph{ROSAT}/PSPC. This choice was motivated by the low background of this instrument, which allowed us to extract high-quality surface-brightness profiles out to large radii \citep{vikhlinin99,eckert11}. We reduced the data using the ESAS software \citep{snowden94}. The non-X-ray background was modeled and subtracted from the observed image, and a vignetting correction was applied. The surface-brightness profile was then extracted from the corrected image in the $0.4-2$ keV band. For the details of the analysis procedure, we refer to \cite{eckert11}.

In this section, we aim at comparing the results of the R-L deprojection method with those obtained with the O-P method. For this purpose, we extracted the emissivity, the density profile, and the temperature profile from the X-ray surface brightness using both methods.

\subsection{Testing the R-L deprojection method}\label{sec:SphericalSym}

The R-L algorithm is an iterative method that tends to deproject the features of the observed projected function on all scales. To avoid reproducing the instrumental fluctuations and other sources of noise in the deprojected profile, Lucy \citep{RL_lucy94} suggested to add a regularization term to a likelihood function whose scale and amplitude were controlled by two parameters, the regularization parameter $\alpha$ and the smoothing scale $L$ \citep[see][for more details]{RL_method}. In the following, we adapt these parameters to the X-ray data obtained by traditional techniques.


\subsubsection{Three-dimensional emissivity}

The mean value and the error bars of the recovered emissivity were obtained using a Monte Carlo (MC) method based on the R-L deprojection method as follows: The value of the surface brightness profile was randomized within each radial ring, according to a Gaussian distribution whose mean was the observed value and whose width was given by the error bars. After $10000$ simulations, the deprojected value and its error are given by the mean and 1$\sigma$ deviation of the resulting distribution. 

In Fig.~\ref{fig:A1689_JX}, we show the results of the X-ray surface-brightness deprojection using the O-P and R-L methods. To obtain these results, the parameters of the R-L method were set to $\alpha=0.01$ and $L=300\,\mathrm{kpc}$. These parameter values are lower than the expectations \citep[see][for more details]{RL_method} and imply that the regulation term introduced in the R-L method is not very effective in the case of Abell~1689. 

More detailed investigations with simulated clusters have shown that the results of the reconstruction depend very little on the values chosen for the regularization parameters. Testing wide ranges for both $\alpha$ and $L$, typically covering an order of magnitude, changes the recovered potentials routinely by much less than the statistical uncertainties.

We emphasize that the regularization term is controlled by the difference of the potential reconstructions between subsequent iterations steps, which vanishes progressively as the iteration proceeds. The regularization is thus gradually switched off toward the end of the iteration. We fixed the number of iterations to 10, which corresponds to the number of iterations at which the algorithm converges, because no significant changes in the emissivity profile were observed when more iterations were used. 

This is endorsed by our result, which shows that the emissivity profiles obtained with the two methods agree very well within the error bars. However, the outermost points of the profile are different in the two techniques, which can be attributed to edge effects \citep[see, e.g.,][]{mclaughlin99}. The large error bars in the outermost points are caused by the weakness of the signal at large radii.

\subsubsection{Three-dimensional density profile}

The temperature of the cluster Abell 1689 lies between $2\,\mathrm{keV}$ and $10\,\mathrm{keV}$ \citep{kawaharada10}. This means that the soft energy band $[0.4;2]\,\mathrm{keV}$ is below the bremsstrahlung cutoff energy, and  the emissivity only  weakly depends on the temperature. In this particular case, the density profile can be expressed as a function of the emissivity, $ j_{[0.4;2]}$, as
\begin{equation}\label{eq:rho}
  \rho(r)\propto \sqrt{j_{[0.4;2]}(r)}\;.
\end{equation} 
In Fig.~\ref{fig:A1689_density} we compare the density profile obtained using Eq.~[\ref{eq:rho}] for the two deprojection methods: the O-P method \citep[in blue,][]{eckert12} and the R-L method (in red, this analysis). Again, the density profile is well reproduced using the R-L procedure. 

\subsubsection{Temperature profile}

Using the polytropic relation (Eq.~[\ref{eq:eqPoly}]) and the fact that we considered only the soft energy band [0.4;2] keV, the temperature profile can be expressed as
\begin{equation}\label{eq:T}
  T(r)\propto \rho(r)^{\gamma-1}\propto j_{[0.4;2]}(r)^{(\gamma-1)/2}\;.
\end{equation} 
Figure~\ref{fig:A1689_Tprofile} shows the temperature profile reconstructed with a polytropic index of $\gamma=1.19\pm0.04$ and the temperature profiles obtained using different techniques. 
 The polytropic index $\gamma=1.19\pm0.04$ results from fitting the observed pressure profile \citep{planck13} with the reconstructed density profile (obtained in Sect.~3.1.2), using Eq.~[\ref{eq:eqPoly}]. 

For completeness, we also fitted Eq.~[\ref{eq:T}] to the joint \textit{XMM-Newton} and \textit{Suzaku} data and obtained $\gamma=1.17\pm0.03$ for the polytropic index with 1$\sigma$ error, which is consistent with our earlier result.
 
 The decrease in the temperature profile supports the validity of the polytropic assumption (Eq.~[\ref{eq:eqPoly}]).
 
\begin{figure}
  \includegraphics[height=\columnwidth,angle=270]{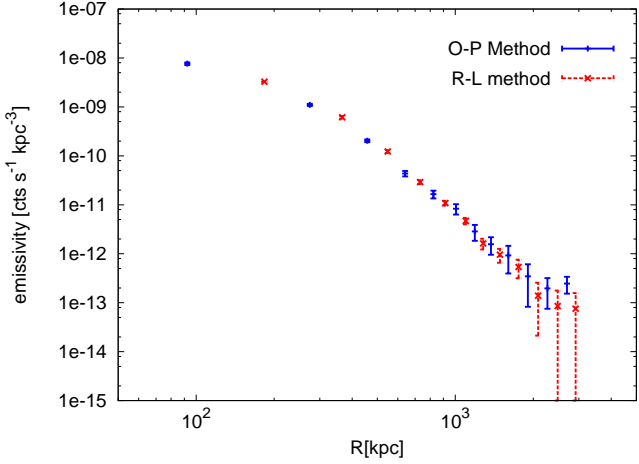}
\caption{Emissivity obtained by applying the R-L (this analysis) and the O-P methods \citep{eckert12} to the ROSAT surface-brightness profile. The reconstructed emissivity using the R-L method has been normalized to the O-P data points within a spherical shell delimited by the radii in the range [183; 2700] kpc.}
\label{fig:A1689_JX}
\end{figure}

\begin{figure}
  \includegraphics[height=\columnwidth,angle=270]{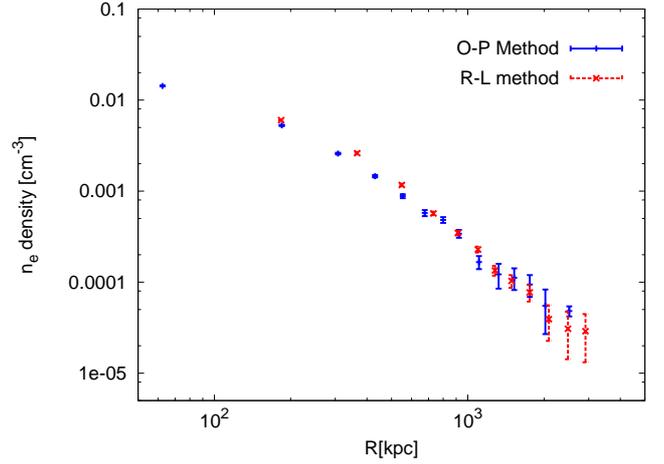}
\caption{Electron-density reconstructed profile obtained from the R-L deprojection method (this analysis) compared to the electron-density profile recovered from the O-P deprojection method \citep{eckert12}.  For both deprojection methods, Eq.~[\ref{eq:rho}] has been used. The reconstructed density using the R-L method has been normalized to the O-P data points within a spherical shell delimited by the radii in the range [183; 2515] kpc.}
\label{fig:A1689_density}
\end{figure}

\begin{figure}
  \includegraphics[height=\columnwidth,angle=270]{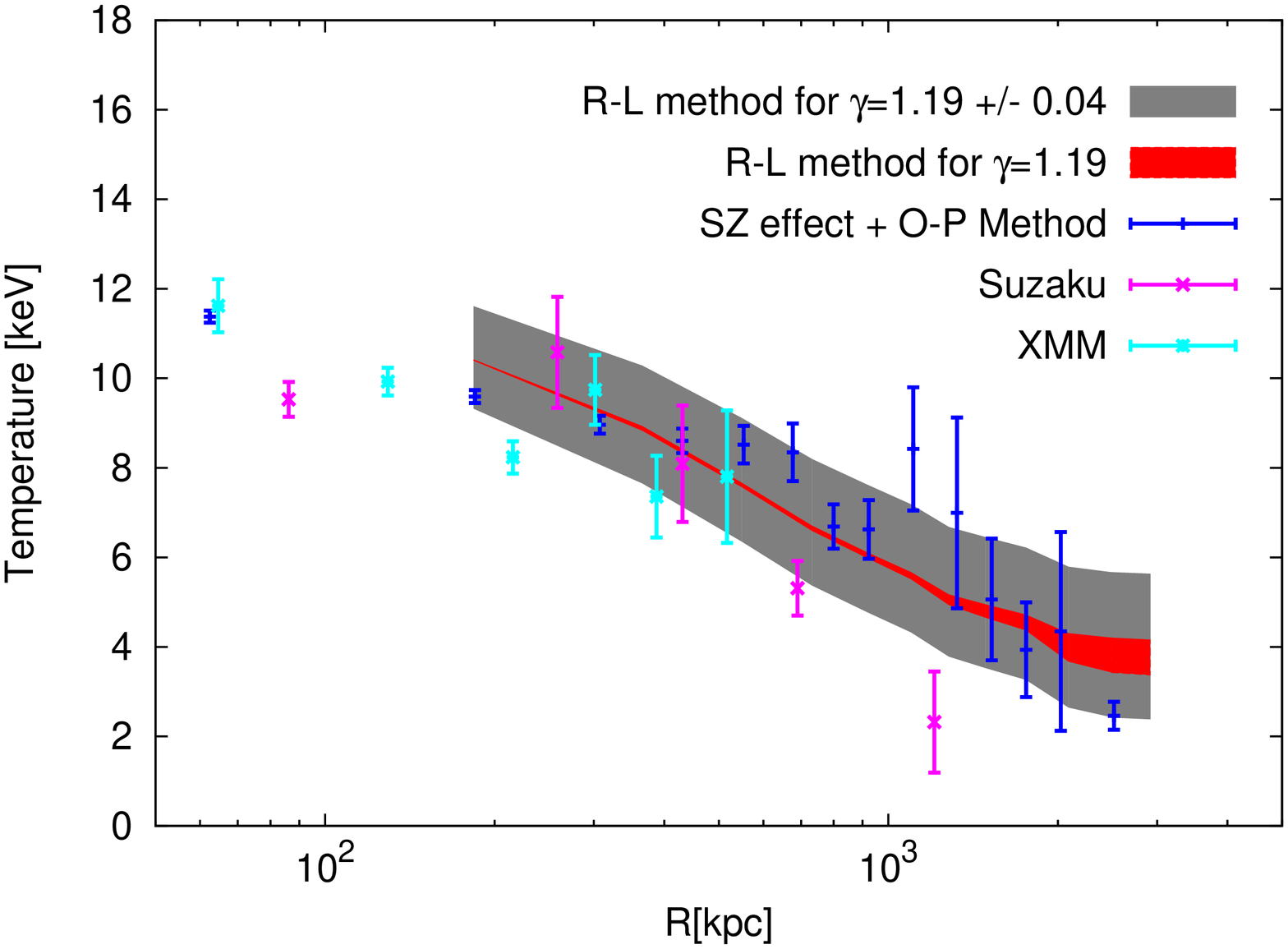}
\caption{Temperature profile obtained from the R-L reconstructed emissivity profile via Eq.~[\ref{eq:T}] with  $\gamma=1.19\pm 0.04$, compared to temperature profiles obtained using different techniques. Dark blue: profile derived from combined X-ray and thermal-SZ measurements \citep{planck13} using the ideal gas assumption \citep{eckert13}; light blue: spectral fitting of XMM observations \citep{snowden08}; pink:  spectral fitting of Suzaku observations \citep{kawaharada10}; red: mean value and errors on the reconstructed profile for $\gamma=1.19$; gray: mean value and errors on the reconstructed profile for $\gamma=1.19\pm 0.04$. The mean and the error bars were obtained with a Monte Carlo method based on the R-L deprojection. The reconstructed temperature profile has been normalized to the data points from \citet{eckert13} within a spherical shell delimited by the radii in the range [183; 2515] kpc.}
\label{fig:A1689_Tprofile}
\end{figure}

\subsection{Advantages of the R-L method compared to the O-P method}\label{sec:RL_OP}

In observations, the only information available is integrated along the line of sight. In the framework of this study, we define as  local a deprojection method that allows independently deprojecting this projected information for each line of sight, meaning that it is not necessary to know the value of the deprojected quantity in one line of sight to derive the deprojected value in another and that regions of different projected radius are not mixed during the deprojection procedure. 

In Sect.~\ref{sec:SphericalSym}, we saw that the R-L and O-P deprojection methods lead to consistent results. We now explain why we aim at using the R-L method instead of the O-P procedure.

At this point of the analysis, it is necessary to briefly review
the concept of the O-P method. We refer to \citet{kriss83} for more details. The O-P method is a geometrical deprojection technique initially introduced using the assumption of spherical symmetry \citep{kriss83}, but it can be generalized to clusters with axial symmetry \citep[see, e.g.,][]{buote12a}. This method assumes that the cluster has an onion-like shell structure, with a uniform emissivity in each of the concentric spherical shells. The three-dimensional profile of the emissivity is then recovered from the surface brightness map in an iterative way from ring to ring, progressing from the outskirts to the center of the cluster.

To carry out this iterative procedure, the O-P method requires the errors to propagate from ring to ring as the observed projected profile is generated. When a quantity is overestimated in a radial bin, the same quantity needs to be underestimated in an adjacent bin, causing the method to produce fluctuations in the deprojected profile \citep[see, e.g.,][]{ameglio07}. Thus, the errors associated with each bin are correlated. The statistical uncertainties associated with the reconstructed three-dimensional profiles are usually estimated using Monte Carlo simulations \citep[see, e.g., ][]{buote00}, which by the same means smoothes the systematic errors due to the fluctuations in adjacent bins.

In the R-L method, the error propagation and the fluctuations due to the deprojection of the instrumental noise are controlled by a regularization term \citep{RL_lucy94}. The choice of this regularization term is adapted to the data themselves and vanishes as the deprojection converges. Furthermore, simulations show that large-scale features are reproduced quickly, while small-scale structures are recovered slowly \citep{reblinsky00,puchwein06}. Therefore,  the R-L method allows controlling the fluctuations that arise from deprojecting the instrumental noise \citep{RL_lucy94}.

An important difference between the two methods is that the O-P method is nonlocal in the sense introduced above: to reconstruct the three-dimensional shell at a given radius, the results of the deprojection of all shells at larger radii are needed, and regions corresponding to different projected radii are mixed. The R-L deprojection method, however, is local in the sense that the two-dimensional profile is independently deprojected along each line of sight \citep[see, e.g.,][]{reblinsky00, puchwein06}, such that the deprojected function is independently evaluated at each image point and that this algorithm can be applied to data sets with incomplete coverage. However, the relations between different projected radii are controlled by the symmetry assumptions, and the information contained in different lines of sight may be mixed, even if the deprojection procedure is local. For instance, assuming spherical symmetry, the information collected at same projected radius is mixed for any line of sight. 

Nevertheless, as for the O-P method, it is possible to generalize the R-L deprojection method to intrinsically spheroiral cluster bodies \citep[see, e.g, ][]{reblinsky00, puchwein06}. In this formalism, the intrinsic shape of the projected body can be taken into account by a suitable (de)projection kernel in the R-L method. As the mixing between regions of different projected radius is a consequence of the symmetries assumption and not of the deprojection procedure, testing different cluster symmetries should help us understand the effect of these mixings on the resulting profiles.

This feature of local deprojection is most relevant for our study since the information provided by gravitational lensing measurements that we use here is  also local. In the SaWLens method developed by \citet{merten09}, estimates of the lensing potential are obtained locally in all cells of a grid covering the region considered on the sky, and therefore, the observations of weak gravitational lensing give information on the local galaxy density. 

\subsection{Reconstructing the lensing potential}\label{test_reconstruction}

The reconstructed projected potential $\psi$ is the projection along the line of sight of the gravitational potential $\phi$. In our formalism, the gravitational potential is expressed in terms of the reconstructed emissivity by \citep[see][for the derivation]{RL_method} 
\begin{equation}\label{eq:phi}
  \phi(r)\propto j_x(r)^{\eta}\;,\quad\eta=\frac{2(\gamma-1)}{3+\gamma}\;.
\end{equation}
The three-dimensional gravitational potential reconstructed this way is then simply projected along the line of sight. In this equation, we use the bolometric emissivity ($j_x$) and not the emissivity restricted to the energy band [0.4;2] keV. To recover the bolometric emissivity from the energy-restricted emissivity (see Fig.~\ref{fig:A1689_JX}), we used the temperature profile (see Fig.~\ref{fig:A1689_Tprofile}) and the thin-plasma code APEC \citep[][fixing the metallicity to 0.3]{smith01} to compute the ratio between the bolometric flux and the restricted flux for each radius. This factor was then used to recover the bolometric emissivity from the restricted emissivity.

The lensing potential is recovered from observations of the distorted images of distant background galaxies. The distortion (shear and magnification) of these images is related to linear combinations of second derivatives of the lensing potential of a foreground cluster. This distortion is due to the gravitational tidal field of a deep potential well between the observer and the background galaxies.  In the case of weak-lensing measurements, the distortion of individual galaxies is too weak to be observed, but the distortion averaged over sufficiently many neighboring galaxies is used to estimate the gravitational tidal field of the foreground galaxy cluster. In a large enough sample, the random intrinsic orientation of the background galaxies is expected to average out, and the ellipticity induced by lensing can be measured. The lensing potential data used here were recovered with SaWLens in a nonparametric way from 
weak-lensing data.

\begin{figure}
  \includegraphics[height=\columnwidth,angle=270]{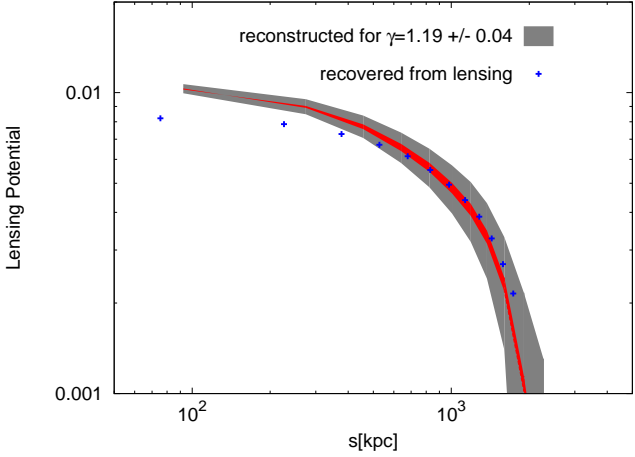}
\caption{Projected gravitational potential reconstructed using the method described in \citet{RL_method} compared to the potential recovered from weak gravitational lensing as a function of the projected radius $s$. The projected potential reconstructed from the X-ray emission has been obtained assuming a polytropic index $\gamma=1.19\pm0.04$ (see the text for details). The potential recovered by lensing is shown in blue; the mean projected potential obtained from X-ray observations with $\gamma=1.19\pm0.04$ is shown in red, the uncertainties of the latter are shown in gray. The mean and errors have been obtained using a Monte Carlo method to randomize the algorithm described in \citet{RL_method}. The reconstructed lensing potential has been normalized to the lensing data within a circular shell region delimited by the radii in the range [92; 1734] kpc.}
\label{fig:A1689_LensingPot}
\end{figure}

In Fig.~\ref{fig:A1689_LensingPot}, we show the reconstructed, two-dimensional potential assuming a polytropic index $\gamma=1.19\pm0.04$. This profile is compared with the lensing potential obtained from weak lensing. At radii $\gtrsim 500\mathrm{~ kpc}$ our reconstructed potential profile agrees well with the reconstructed lensing profile within the error bars, even though the potential reconstructed from the X-ray emission appears to be slightly less curved. The discrepancy at small radii and the slight curvature change at larger radii may be caused by the lack of resolution in the weak-lensing measurement and by the other assumptions made (sphericity, polytropic and equilibrium assumptions). These points are discussed in more
detail in the next section.

As we have mentioned in the introduction, studying the intracluster gas may help us understand the dark-matter distribution in the cluster because the ICM fills the dark-matter potential well. As a result of the locality of the R-L deprojection, our method opens a way for cluster potentials to complement the information provided by lensing and joins two entirely different observables on the grounds of the gravitational potential, which provides a new tool for constraining the physical state of the gas locally.

\section{Discussion}\label{sec:discussion}

We used the R-L deprojection method to recover the three-dimensional gas distribution in Abell~1689 from its X-ray surface brightness. We then converted this gas distribution into a three-dimensional gravitational potential that we subsequently projected for comparison with the lensing potential. Using the R-L deprojection method is advantageous because it allows independently deprojecting the X-ray observables along each line of sight as well as treating data sets with incomplete coverage. This therefore allows excluding regions where the equilibrium assumptions may not hold. As a result, a substantial amount of information becomes available by the direct comparison of the lensing potential with the projected potential reconstructed from X-ray emission (Fig.~\ref{fig:A1689_LensingPot}). 

For example, the agreement between the two reconstructed potentials at radii from $500-2000\,\mathrm{~kpc}$ indicates that the assumptions underlying the derivation of the projected potential from X-ray observations, namely~hydrostatic equilibrium, polytropic stratification, and the ideal-gas equation of state [Eqs.~(\ref{eq:eqHydro})-(\ref{eq:Brem})] are not substantially violated in the cluster Abell~1689 within this radial range. Nevertheless, we can observe a slight discrepancy between the potentials curvature: the potential recovered from lensing is flatter than the one reconstructed from the X-ray observations. This flattening can be produced by the large resolution limit  present in the weak-lensing measurements \citep[$\sim 100$~kpc, see, e.g., ][for a review]{bartelmann03}, which can smooth the data so that the lensing potential recovered from weak-lensing measurements appears flatter than the potential recovered from X-ray observations. This discrepancy is expected to vanish if more potential observables are taken into account. For instance, a combined strong- and weak-lensing reconstruction is expected to yield a better match in the region close to the cluster center because the strong lensing is sensitive to this region.
\paragraph{}
At small radii ($R\lesssim 500\mathrm{~kpc}$), however, the two projected potentials differ. Such a discrepancy in the central region of the cluster was also found by \citet{peng09}, \citet{andersson04}, \citet{morandi11}, \citet{miralda95}, \citet{lokas06}, and \citet{sereno13}, as mentioned in the introduction. This could be explained by a modification of the balance between the physical processes in the inner part of the cluster. It may indicate that some of the assumptions made should not be used in the central part of the cluster. 

For instance, the polytropic assumption with constant polytropic index may not be valid in the central part of cool-core clusters because the radiative cooling near the center renders the cooling time shorter than the Hubble time \citep[e.g., ][]{fabian94}. Cooling alters the polytropic equation by lowering the temperature of the gas, which implies that the temperature profile does not increase toward the center as the gas density does, but instead decreases. Therefore, the assumption of a constant polytropic index is not expected to be correct at the cluster centre. However, the assumption that the gas follows the polytropic equation seems to be valid in Abell 1689, with a fitted polytropic index of $\gamma=1.19\pm 0.04$ (Fig.~\ref{fig:A1689_Tprofile}).

On the other hand, the assumption of hydrostatic equilibrium is likely valid in the central regions of cool-core clusters because of the deep gravitational potential \citep[see, for instance, Fig.~2 of ][]{lau09}. 
This may not be the case in non-cool-core clusters. As has been shown by \citet{mahdavi13}, for instance, departure from the equilibrium assumption is expected in unrelaxed clusters.
The authors compared the lensing mass with the mass obtained under hydrostatic equilibrium assumptions for a sample of clusters at $R\lesssim R_{500}$ ($\sim1.34\mathrm{~Mpc}$ in the case of Abell~1689) and concluded that the departure from the equilibrium assumption is different between relaxed and merging clusters: while for cool-core clusters, almost no bias is observed between the lensing and the hydrostatic mass, a bias of 15-20\% is present for non-cool-core clusters. Abell~1689 is known to experience a merging event aligned with the line of sight, and therefore, the gas is not expected to be in hydrostatic equilibrium near the cluster center \citep[see also, e.g.,][]{nelson12, morandi11}. This may contribute to the discrepancy at small radii observed in Fig.~\ref{fig:A1689_LensingPot}.

Furthermore, any departure from spherical symmetry could bias the estimate of the cluster mass, and if where the major axis of the cluster is preferentially aligned with the line of sight, it should lead to an overestimation of the lensing mass with respect to the X-ray mass. This is consistent with the conclusions of  \citet{lee03,lee04}, who showed that the dark matter profile in nonspherical clusters is always more sensitive to triaxiality than the gas distribution from X-ray measurements. This weak dependence of the X-ray measurements on the cluster shape was also observed in \citet{buote12a, buote12b}, where the authors computed  the biases in the observables obtained by applying a spherical O-P deprojection method on clusters with significant triaxiality. They concluded that spherical averaging biases the observables by a small factor ($<1\,\%$) with respect to a triaxial deprojection. 

Therefore, the nonsphericity of the cluster could explain the discrepancy at small radii between the X-ray and lensing projected potential profiles for Abell
1689, but interestingly, it does not seem to strongly affect the reconstruction of the potential in the region $500-2000\,\mathrm{~kpc}$. 

\paragraph{}
At large radii ($R\gtrsim R_{500}$), simulations tend to show that the equilibrium assumptions are not valid any more due to mixing of the ICM with the infalling material from the large-scale structure \citep[see, e.g,][for a review]{reiprich13}. The deviation from the hydrostatic equilibrium can be due to residual gas motions or incomplete thermalization of the ICM \citep{mahdavi13}. Such phenomena are expected in the outskirts of clusters, where matter is accreted and substructures such as clumps are more likely to be found \citep{nagai11,vazza13,zhuravleva13}, so that our reconstruction method is not expected to be valid at radii larger than $R_{500} (\sim 1.339\mathrm{~Mpc}$, for Abell 1689). However, as the lensing data used in our study do not extend to much larger radii than $R_{500}$, we cannot test this statement. When we
extrapolated the lensing data points to larger radii, we did not find significant differences between X-ray and lensing information in this radial range.

\paragraph{}
Some systematic effects arise from the different assumptions we made in treating the X-ray observations. In Eq.~[\ref{eq:Brem}], we assumed that the emissivity is dominated by bremsstrahlung emission. This approximation represents a difference of at most 8\% in the normalization compared to the normalization we would obtain taking into account the contribution of the emission lines for a metallicity of 0.3. Then, the conversion from 0.4-2~keV to the bolometric emissivity, computed assuming a temperature profile that follows Eq.~[\ref{eq:T}] with $\gamma=1.19$, implies systematic errors of about 4\% at most. The deviations from a flat field with the ROSAT satellite have been computed to introduce systematic effects on the order of  6\% of the background in \citet{eckert12}. This value corresponds to the scatter in the residual, resulting from the fit of the surface brightness profiles of flat fields with a constant value. Finally, we estimated the effect of the limited ROSAT PSF on the reconstructed profile by simulating a beta model with parameters consistent with those of Abell 1689 and by convolving it with the PSF of ROSAT. While the convolved profile is significantly flatter than the simulated profile inside $\sim$300~kpc (by $\sim18\%$), beyond this radius the effect of the PSF is weak, and we estimated it to be at the level of $\sim3\%$ of the observed signal. In general, these instrumental effects are not expected to change our conclusions qualitatively. Furthermore, a combination of other cluster observables (such as SZ signal or galaxy kinematics) is expected to constrain the effects of these systematics and to help us build a consistent model for the projected gravitational potential.

\section{Conclusion}\label{sec:conclusion}

We applied a newly developed method \citep{RL_method} for reconstructing the projected gravitational potential of the well-known galaxy cluster Abell~1689. We compared the X-ray profiles obtained using the onion-peeling deprojection technique with the profiles  from Richardson-Lucy deprojection. Figures~\ref{fig:A1689_JX}-\ref{fig:A1689_Tprofile} show the bremsstrahlung emissivity, electron density, and temperature profiles obtained using the two deprojection methods. All profiles are consistent within the error bars.

Figure~\ref{fig:A1689_LensingPot} shows the projected gravitational potential reconstructed from X-ray data following the procedure described in \citet{RL_method}. This figure shows that the projected potential obtained directly from X-ray observations agrees with the lensing potential obtained from weak-lensing measurements. 

Therefore, for sufficiently precise data sets, the R-L deprojection method is expected to help us constrain the validity of the assumed equilibrium assumptions at all projected radii.
We obtained these results using a spherical symmetry assumption. However, a possible bias coming from this assumption can also contribute to the discrepancy between the X-ray and lensing potentials at small radii (Fig.~4). To remove the degeneracy between the different explanations (breakdown of hydrostatic equilibrium or projection effects), we are now working on generalizing this deprojection method using a triaxial kernel. 

\paragraph{}
We provided a general proof of concept of potential reconstruction combining X-ray data with weak-lensing measurements. The next step now consists of combining all observables provided by galaxy clusters: strong and weak gravitational lensing, X-ray surface-brightness and temperature, galaxy kinematics, and the thermal Sunyaev-Zel'dovich effect into a joint construction of one consistent model for the projected cluster potential. This combination will provide a substantial amount of information and will help us to quantify the validity of the assumptions made. If the assumptions are valid, the combination of all observables will produce the best constrained galaxy-cluster potentials.

\acknowledgements{We are grateful to Julian Merten for his many valuable insights and for the lensing data he provided for this analysis, and to Sara Konrad for her contribution in developing this reconstruction method. This project was supported in part by the Baden-W\"urttemberg Foundation under project ``Internationale Spitzenforschung II/2'' and by the Collaborative Research Area TR-33 ``The Dark Universe'' as well as project BA 1369/17 of the Deutsche Forschungsgemeinschaft and in part by the Swiss National Science Foundation.}

\end{document}